%% file: sn-article.tex
\documentclass[pdflatex,sn-mathphys-num]{sn-jnl}

\usepackage{graphicx}%
\usepackage{multirow}%
\usepackage{amsmath,amssymb,amsfonts}%
\usepackage{amsthm}%
\usepackage{mathrsfs}%
\usepackage[title]{appendix}%
\usepackage{xcolor}%
\usepackage{textcomp}%
\usepackage{manyfoot}%
\usepackage{booktabs}%
\usepackage{algorithm}%
\usepackage{algorithmicx}%
\usepackage{algpseudocode}%
\usepackage{listings}%

\theoremstyle{thmstyleone}%
\theoremstyle{thmstyletwo}%

\theoremstyle{thmstylethree}%

\begin{document}

\title[Constructing the Padmanabhan Holographic Model in a BIonic System]{Constructing the Padmanabhan Holographic Model in a BIonic System}

\author[1]{\fnm{Alireza} \sur{Sepehri}}\email{dralireza.sepehri14@gmail.com} 

\author[2]{\fnm{Muhammad Al-Zafar} \sur{Khan}}\email{Muhammad.Khan@zu.ac.ae} 
\equalcont{These authors contributed equally to this work.}

\affil[1]{\orgname{Sepehr Mohaghegh Kosar}, \orgaddress{\city{Mashhad}, \country{Iran}}}

\affil[2]{\orgdiv{College of Interdisciplinary Studies}, \orgname{Zayed University}, \orgaddress{\city{Abu Dhabi}, \country{UAE}}}

\abstract{Recently, Padmanabhan has argued that a difference between the number of degrees of freedom on the surface and the number in a bulk causes the expansion of the universe. We can reconsider this idea in a BIon system. A Bion is formed from two branes that are connected by a wormhole. Our universe may live on one of these branes.  Each brane could be formed by joining lower-dimensional branes, such as $D_1$ ones. By joining $D_1$ branes, a $D_n$ brane is formed, and some amounts of energy are released. Then, perhaps some dimensions are compacted, and a certain amount of energy is released. These energies cause a significant difference between the number of degrees of freedom on the surface and in the bulk of branes. This causes the evolution of the universe and many changes in thermodynamic parameters, such as entropy, and cosmic parameters, including the Hubble constant. We obtain the standard form of the Hubble parameter and its dependency on redshift in a Bion system.}

\keywords{Brane Universe, Hubble Parameter, Extra Dimensions}



\maketitle

\section{Introduction}\label{intro}
Some years ago, Padmanabhan proposed a model that explains the evolution of the universe from the perspective of an observer looking from outside it. In this model, the accelerating expansion of the universe is due to the difference between the number of degrees of freedom on the surface of the universe and the number of degrees of freedom in the bulk. This difference in the number of degrees of freedom has a direct impact on cosmological variables such as the Hubble parameter and entropy \cite{padmanabhan2012emergence}. Padmanabhan's idea has been widely used in cosmological and gravitational studies. For example, using this idea, Friedman equations and other evolution equations of the universe have been obtained not only in four dimensions and Einstein gravity but also in higher dimensions and other gravity theories like the Lovelock and Gauss-Bonnet ones \cite{padmanabhan2012emergence,yang2012emergence,ling2013note,sheykhi2013friedmann,eune2013emergent,chang2014friedmann}. Also, some authors have used Padmanabhan idea in GUP \cite{ali2014emergence}. 

But the basic question is, what is the real concept of the universe surface and the bulk in the Padmanabhan model? In response to this question, some authors have used the Padmanabhan model in the BIon system. The BIon consists of two branes connected by a wormhole. Two universes are born on each of the branes. By injecting energy through the wormhole into each of the branes, the universe associated with them expands. This energy has a relation to the difference in the number of degrees of freedom on the surface of the brane and the number of degrees of freedom in the bulk. In this model, the bulk includes the wormhole and additional dimensions perpendicular to the brane \cite{sepehri2015emergence,beesham2018emergence,sepehri2016emergence,sepehri2015emergence2,grignani2017flowing,harmark2000supergravity,ghaforyan2018tsallis}. Wormholes could be a $D_1$ or $D_2$ brane. By dissolving $D_1$ branes into universe branes, they expand. On the other hand, universe branes could form from joining $D_1$ branes. During the formation of $D_n$ branes from $D_1$ branes, some amount of energy is produced. This energy could be another cause of the acceleration of the universe's expansion \cite{sepehri2015emergence,beesham2018emergence,sepehri2016emergence,sepehri2015emergence2}. In this paper, we obtain the relation between the Hubble parameter and the BIon coordinates.  

This paper is divided as follows:
\begin{itemize}
\item In Sec. \ref{the hubble parameter changes} we discuss how the Hubble parameter undergoes a change in the BIon system.
\item In Sec. \ref{sec:sec2} we discuss the effects of the released energy on the Hubble parameter when the number of dimensions and compactifications are increased.
\item In Sec. \ref{conclusion}, we provide concluding remarks whereupon the results of this research are contemplated, and closing arguments are furnished. 
\end{itemize}
 
\section{The Hubble Parameter Changes in the BIon System}\label{the hubble parameter changes}

Recently, some authors have shown that our universe could have been born on one of the BIon's branes. A BIon is a system consisting of two branes and a wormhole. The branes exchange energy with each other through the wormhole. The metric of this BIon in ten-dimensional spacetime is written as follows \cite{grignani2017flowing,harmark2000supergravity}
\begin{equation}\label{eqn:metric}
ds^{2}=ds_{yz}^{2}+ds_{tx}^{2}+ds_{ru}^{2},
\end{equation}
where the elements of the metric \eqref{eqn:metric} could be introduced as
\begin{equation}\label{eqn:elements of metric}
\left\{
\begin{aligned}
ds_{yz}^{2}=&\;B^{-1/2}C^{-1/2}\left(dy^{2}+dz^{2}\right), \\
ds_{tx}^{2}=&\;B^{1/2}C^{-1/2}\left(-f\;dt^{2}+dx^{2}\right), \\
ds_{ru}^{2}=&\; B^{-1/2}C^{1/2}\left(f^{-1}dr^{2}+r^{2}du_{5}^{2}\right),
\end{aligned}
\right.
\end{equation}
with the parameters in \eqref{eqn:elements of metric} defined as \cite{sepehri2015emergence,beesham2018emergence,sepehri2016emergence,sepehri2015emergence2,grignani2017flowing,harmark2000supergravity,ghaforyan2018tsallis}
\begin{equation}\label{eqn:parameters of metric}
\left\{
\begin{aligned}
f=&\;1-\left(r_{0}r^{-1}\right)^{-4}, \\
C=&\;1+\left(r_{0}r^{-1}\right)^{4}\sinh^{2}b, \\
B=&\;\cos^{2}\tilde{n}+C^{-1}\sin^{2}\tilde{n}, \\
\cosh^{2}b=&\;\text{Cs}^{1}_{q}+\text{Cs}^{2}_{q}, \\
\text{Cs}^{1}_{q}=&\;\frac{3}{2}\cos^{-1}(q)\cos(q/3), \\
\text{Cs}^{2}_{q}=&\;\frac{3\sqrt{3}}{2}\sin(q/3)\cos^{-1}(q), \\
\cos\tilde{n}=&\;\left(1+k^{2}r^{-4}\right)^{-1/2}, \\
\cos q=&\;4.3\left(T_{D_{3}}N\right)^{1/2}T^{4}\left(1+k^{2}r^{-4}\right)^{1/2}.
\end{aligned}
\right.
\end{equation}
In \eqref{eqn:parameters of metric}, $T$ denotes the temperature, $T_{D_{3}}$ the tension of the brane, $N$ is the number of branes, and $k$ is a constant. The above equation shows that the metric of a BIon depends on the temperature, the size of the BIon, and its distance from other BIon. The separation distance between two branes along the $z$-coordinate in a BIonic system is obtained from the following equation
\begin{equation}\label{eqn:bion distances}
d_{r}z=\left[F^{2}(r)F^{-2}(r_{0})-1\right]^{-1/2},    
\end{equation}
where 
\begin{equation}
F(r)=r^{2}\cosh^{-4}b\;\left(4\cosh^{2}b-3\right),
\end{equation}
with $d_{r}$ in \eqref{eqn:bion distances} denotes the derivative with respect to $r$. Additionally, $r_{0}$ is related to the size of the branes at the moment of their collisions and the complete injection of the wormhole energy into one or both of the branes. The equation above \eqref{eqn:bion distances} shows that the separation distance between two branes depends on the temperature and size of the branes. 

On the other hand, the mass of the BIon changes as the distance between the branes changes. The following equation examines the rate of mass change depending on the temperature and size of the branes
\begin{equation}\label{eqn:rate of mass change}
d_{z}M_{\text{BIon}}=m_{T}(T)m_{r}(r)m_{b}(b),
\end{equation}
where
\begin{equation}\label{eqn:parameters in rate of mass change}
\left\{
\begin{aligned}
m_{T}(T)=&\;1.7^{-1}\left(T_{D_{3}}T^{-2}\right)^{2}, \\
m_{r}(r)=&\;r^{2}F(r)F^{-1}(r_{0}), \\
m_{b}(b)=&\;\cosh^{-4}b\;\left(4\cosh^{2}b+1\right).
\end{aligned}
\right.
\end{equation}

Entropy also depends on the temperature of the bion and the size of the branes. This dependence is examined by the following equation
\begin{equation}\label{eqn:entropy}
d_{r}S_{\text{BIon}}=S_{T}(T)S_{r}(r)S_{b}(b),
\end{equation}
where
\begin{equation}\label{eqn:relations in entropy}
\left\{
\begin{aligned}
S_{T}(T)=&\;\left(6.8L_{p}\right)^{-1}\left(T_{D_{3}}T^{-2}\right)^{2},\\
S_{r}(r)=&\;r^{2}F(r)\;\left[F^{2}(r)-F^{2}(r_{0})\right]^{-1/2}, \\
S_{b}(b)=&\;\cosh^{-4}b.
\end{aligned}
\right.
\end{equation}

According to the Padmanabhan model, a relationship exists between the number of degrees of freedom of the universe and changes in mass and entropy. For example, the derivative of the sum of the degrees of freedom on the surface of the universe and the degrees of freedom around it in bulk respect to brane coordinates is related to the entropy of the universe, while the derivative of the difference of these degrees of freedom is related to the changes in mass on the $z$-axis. This is expressed as
\begin{equation}\label{eqn:bulk and surface derivatives}
\left\{
\begin{aligned}
d_{r}\left(N_{\text{sur}}+N_{\text{bulk}}\right)=&\;\left(4L_{p}\right)^{2}d_{r}S_{\text{BIon}}, \\
d_{r}\left(N_{\text{sur}}-N_{\text{bulk}}\right)=&\;d_{z}M_{\text{BIon}}\times d_{r}z,
\end{aligned}
\right.
\end{equation}
where $N_{\text{sur}}$ and $N_{\text{bulk}}$ are the number of degrees of freedom on the surface and bulk, respectively. Using \eqref{eqn:bulk and surface derivatives}, we obtain
\begin{equation}\label{eqn:derivative of surface}
d_{r}N_{\text{sur}}=\frac{1}{2}\left[\left(4L_{p}\right)^{2}d_{r}S_{BIon}+d_{z}M_{\text{BIon}}\times d_{r}z\right].
\end{equation}
On the other hand, the number of degrees of freedom on the surface depends on the Hubble parameter ($H$) and scale factor ($a$) according to
\begin{equation}\label{eqn:relation between DOF of surface and Hubble}
N_{\text{sur}}=12.8L_{p}^{-2}r_{A}^{2},    
\end{equation}
where
\begin{equation}\label{eqn:parameter inside hubble}
r_{A}=\left(H^{2}+K'a\right)^{1/2}. 
\end{equation}
Using the results from equations \eqref{eqn:rate of mass change}-\eqref{eqn:parameter inside hubble}, we obtain the Hubble parameters in terms of time
\begin{equation}\label{eqn:hubble relation in terms of time}
H^{2}=H_{0}^{2}\left[1-G_{0}t^{3}(t-t_{0})^{-3}+\ldots\right]^{2},
\end{equation}
where $t_{0}$ is the time of colliding branes. Equation \eqref{eqn:hubble relation in terms of time} shows that by passing time and closing branes to each other, the Hubble parameter increases and shrinks to infinity. By using the relation between redshift and time
\begin{equation}
Z_{R}=t_{0}t^{-1},
\end{equation}
we can rewrite \eqref{eqn:hubble relation in terms of time} as 
\begin{equation}\label{eqn:hubble parameter in terms of redshift}
H^{2}=H_{0}^{2}\left[g_{1}\left(Z_{R}+1\right)^{3}+\ldots\right].
\end{equation}
Equation \eqref{eqn:hubble parameter in terms of redshift} is similar to the previous predictions for the Hubble parameter. Further, \eqref{eqn:hubble parameter in terms of redshift} shows that by increasing the redshift, the Hubble parameter grows. This model is consistent with observations.


\section{The Effect of the Energy Released from Increasing the Number of Dimensions and the Compactness of Some Dimensions on the Hubble Parameter}\label{sec:sec2}

Previously, it has been shown that any $D_n$ brane can be formed by joining lower-dimensional branes, such as $D_1$ branes.  By joining a $D_1$ brane to a $D_{n–1}$ brane, a $D_n$ brane is formed, and an extra energy becomes free. We can write \cite{sepehri2016emergence,sepehri2015emergence2}
\begin{equation}
E_{D_{n}}=E_{D_{n-1}}+E_{D_{1}}-V_{\text{separation}}.
\end{equation}
Each $D_n$ brane could be formed by joining $n$ $D_1$ branes. We can write \cite{sepehri2016emergence,sepehri2015emergence2}
\begin{equation}
E_{D_{n}}=nE_{D_{1}}-nV_{\text{separation}}.
\end{equation}
On the other hand, compacting some dimensions releases a certain amount of energy. We can write \cite{sepehri2016emergence,sepehri2015emergence2}
\begin{equation}
E_{D_{n}}=E_{D_{n+m}}-P_{n}(m_{n})m_{n}V_{n,\text{compact}},
\end{equation}
where $P_{n}$ is the probability of compactification and $V_{n}$ is the potential of compactification. Summing up all potential energies gives
\begin{equation}
V_{n}=P_{n}(m_{n})m_{n}V_{n,\text{compact}}+nV_{\text{separation}}. 
\end{equation}
This potential has a direct impact on cosmic parameters, including the Hubble parameter, metric, and entropy. For example, thermal Bionic metric changes to
\begin{equation}\label{eqn:generalized metric}
\left\{
\begin{aligned}
d(s_{n})^{2}=&\;d(s_{n-1})^{2}\left[d(s_{n,yz})^{2}+d(s_{n,tx})^{2}+d(s_{n,ru})^{2}\right], \\
d(s_{n,yz})^{2}=&\;d(s_{n-1,yz})^{2}\left[B_{n}^{-1/2}C_{n}^{-1/2}\left(dy_{n}^{2}+dz_{n}^{2}\right)\right], \\
d(s_{n,tx})^{2}=&\;d(s_{n-1,tx})^{2}\left[B_{n}^{1/2}C_{n}^{-1/2}\left(-f_{n}\;dt_{n}^{2}+dx_{n}^{2}\right)\right], \\
d(s_{n,ru})^{2}=&\;d(s_{n-1,ru})^{2}\left[B_{n}^{-1/2}C_{n}^{1/2}\left(f_{n}^{-1}dr_{n}^{2}+r_{n}^{2}du_{n_{5}}^{2}\right)\right], 
\end{aligned}
\right.
\end{equation}
where $n$ is the number of $D_1$ branes that are dissolving into branes and causing their expansions. Additionally, the metric parameters of \eqref{eqn:generalized metric} are
\begin{equation}\label{eqn:generalized functions}
\left\{
\begin{aligned}
f_{n}=&\;1-\left(r_{n_{0}}r_{n}^{-1}\right)^{-4}f_{n-1}, \\
f_{n-1}=&\;1-\left(r_{n-1,0}r_{n-1}^{-1}\right)^{-4}f_{n-2}, \\
\vdots \\
f_{0}=&\;0.
\end{aligned}
\right.
\end{equation}
Equation \eqref{eqn:generalized functions} shows that metric parameters depend on the number of dissolved D1 branes in the BIon system. On the other hand, dissolved $D_1$ branes produce an acceleration that changes the coordinates of the branes and makes the spacetime curved. We have the following relation between acceleration and potential energy of dissolved branes
\begin{equation}
a_{n}=\left(V_{n-1}\right)^{-1}d_{r}V_{n}.
\end{equation}
This acceleration changes the coordinates of branes. We can write
\begin{equation}\label{eqn:generalized acceleration and time}
\left\{
\begin{aligned}
r_{n-1}=&\;a_{n}^{-1}\cosh(a_{n}t_{n})\exp(a_{n}r_{n}), \\
t_{n-1}=&\;a_{n}^{-1}\sinh(a_{n}t_{n})\exp(a_{n}r_{n}).
\end{aligned}
\right.
\end{equation}
Equation \eqref{eqn:generalized acceleration and time} shows that by dissolving a new $D_1$ brane, the coordinates of the brane change again and expand more, and the time becomes more curved. However, at the Big Bang, that acceleration is approximately infinite; space and time coordinates are approximately zero. Then, over time, they expand. We can write
\begin{equation}
t_{0}=0, \quad r_{0}=0. 
\end{equation}

It is concluded that the coincidence of the birth of the universe at the Big Bang, with coordinates zero, or may be compacted to a point. However, as time passes and the $D_1$ branes join, the space and time coordinates expand, and the number of dimensions increases.

By joining $D_1$ branes, the time and space coordinates evolve, causing the metric elements to change. For example, we can write 
\begin{equation}\label{eqn:generalized relations}
\left\{
\begin{aligned}
C_{n}=&\;\left[1+\left(r_{n,0}r_{0}^{-1}\right)^{4}\sinh^{2}b_{n}\right]C_{n-1}, \\
B_{n}=&\;\left(\cos^{2}\tilde{n}_{n}+C_{n-1}\sin^{2}\tilde{n}\right)B_{n-1}, \\ 
\cosh^{2}b_{n}=&\;\text{Cs}^{1}_{q,n}+\text{Cs}^{2}_{q,n}, \\
\text{Cs}^{1}_{q,n}=&\;\frac{3}{2}\text{Cs}^{1}_{q,n-1}\cos^{-1}(q_{n})\cos(q_{n}/3), \\
\text{Cs}^{2}_{q,n}=&\;\frac{3\sqrt{3}}{2}\text{Cs}^{2}_{q,n-1}\sin(q_{n/3})\cos^{-1}(q_{n}), \\
\cos\tilde{n}_{n}=&\;\left(1+k_{n}^{2}r_{n}^{-4}\right)^{-1/2}\cos\tilde{n}_{n-1}, \\
\cos q_{n}=&\;4.3\left(T_{D_{3}}N_{n}\right)^{1/2}T_{n}^{4}\left(1+k_{n}^{2}r_{n}^{-4}\right)^{1/2}\cos q_{n-1}.
\end{aligned}
\right.
\end{equation}

The relations in \eqref{eqn:generalized relations} show that metric parameters depend on the number of dissolved $D_1$ branes and the number of compacted dimensions. Additionally, the temperature of the system depends on these parameters, and by dissolving $D_1$ branes or increasing the compactified dimensions.

The energy of the compactified dimensions and dissolved branes has a direct effect on the separation distance between branes. We can write
\begin{equation}\label{eqn:result1}
d_{r}z_{n}=d_{r}z_{n-1}\left[F_{n}^{2}(r_{n})F_{n}^{-2}(r_{0,n})-1\right]^{-1/2},
\end{equation}
where 
\begin{equation}\label{eqn:result2}
F_{n}(r_{n})=F_{n-1}(r_{n-1})r_{n}^{2}\cosh^{-4}b_{n}\;\left(4\cosh^{2}b_{n}-3\right).
\end{equation}
Equations \eqref{eqn:result1} and \eqref{eqn:result2} show that by joining $D_1$ branes, firstly a BIon is formed and $D_n$ branes are separated from each other. However, eventually, a $D_2$ brane or wormholes that connect $D_n$ branes dissolve into them, and $D_n$ branes move toward each other and finally collide.

The energy of dissolved $D_1$ branes and compactification change the mass of the BIon according to
\begin{equation}\label{eqn:result5}
\begin{aligned}
d_{z}M_{n,\text{BIon}}=&\;d_{z}M_{n-1,\text{BIon}}\left[m_{n,T}(T_{n})+m_{n,r}(r_{n})\right. \\ 
&\left.+m_{n,b}(b_{n})\right],
\end{aligned}
\end{equation}
where
\begin{equation}
\left\{
\begin{aligned}
m_{n,T}(T_{n})=&\;1.7^{-1}\left(T_{D_{3}}T_{n}^{-2}\right)^{2}m_{n-1,T}(T_{n-1}), \\ 
m_{n,r}(r_{n})=&\;r_{n}^{2}F(r_{n})F^{-1}(r_{n_{0}})m_{n-1,r}(r_{n-1}), \\
m_{n,b}(b_{n})=&\;\cosh^{-4}b_{n}\;\left(4\cosh^{2}b_{n}+1\right)m_{n-1,b}(b_{n-1}).
\end{aligned}
\right.
\end{equation}

In fact, by dissolving more $D_1$ branes in BIon, its mass increases. This causes other thermodynamic parameters, like the entropy change. We can write 
\begin{equation}
d_{r}S_{n,\text{BIon}}=d_{r}S_{n-1,\text{BIon}}\left[S_{n,T}(T_{n})S_{n,r}(r_{n})S_{n,b}(b_{n})\right],
\end{equation}
where
\begin{equation}
\left\{
\begin{aligned}
S_{n,T}(T_{n})=&\;S_{n,T}(T_{n-1})\left(6.8L_{p}\right)^{-1}\left(T_{D_{3}}T_{n}^{-2}\right)^{2}, \\
S_{n,r}(r_{n})=&\;S_{n-1,r}(r_{n})F(r_{n})\left[F^{2}(r_{n})-F^{2}(r_{n,0})\right]^{-1/2}, \\
S_{n,b}(b_{n})=&\;\cosh^{-4}b_{n}\;S_{n-1,b}(b_{n}).
\end{aligned}
\right.
\end{equation}

Changes in entropy and mass of BIon cause differences in the number of degrees of freedom on the surface and in the bulk.  We can write
\begin{equation}\label{eqn:result3}
\left\{
\begin{aligned}
d_{r}\left(N_{n,\text{sur}}+N_{n,\text{bulk}}\right)=&\;\left(4L_{p}\right)^{2}d_{r}S_{n,\text{BIon}}, \\
d_{r}\left(N_{n,\text{sur}}-N_{n,\text{bulk}}\right)=&\;d_{z}M_{n,\text{BIon}}\times d_{r}z_{n}.
\end{aligned}
\right.
\end{equation}

Using the relations in \eqref{eqn:result3}, we can obtain the relation between the number of degrees of freedom on the surface and entropy and mass of BIon
\begin{equation}\label{eqn:result4}
d_{r}N_{n,\text{sur}}=\frac{1}{2}\left[\left(4L_{p}\right)^{2}d_{r}S_{n,\text{BIon}}+d_{z}M_{n,\text{BIon}}\times d_{r}z_{n}\right].
\end{equation}

On the other hand, the number of degrees of freedom on the surface has the following relation with the Hubble parameter and the scale factor
\begin{equation}\label{eqn:result6}
\left\{
\begin{aligned}
N_{n,\text{sur}}=&\;12.8L_{p}^{-2}r_{n,A}^{2}, \\
r_{n,A}=&\;\left(H_{n}^{2}+K'a_{n}^{-2}\right)^{1/2}. 
\end{aligned}
\right.
\end{equation}

Solving \eqref{eqn:result4} and using the relations in \eqref{eqn:result5}-\eqref{eqn:result6}, we obtain
\begin{equation}\label{eqn:result7}
H_{n}^{2}=H_{n-1}^{2}\left[1+G_{n}t^{2n+1}\left(t-t_{n}\right)^{-2n-1}+\ldots\right]^{2n}.
\end{equation}

This equation shows that the Hubble parameter depends on the number of dissolved branes and the energy of compacted dimensions. During the dissolving of a $D_1$ brane or the compaction of a dimension, the time coordinates change, and the Hubble parameter increases. 

We can rewrite \eqref{eqn:result7} in terms of red shift. We have 
\begin{equation}\label{eqn:result8}
Z_{R_{n}}=t_{n}t^{-1}
\end{equation}
where $Z_{R_{n}}$ is the red shift during dissolving $n D_1$ branes. Using \eqref{eqn:result8} we can write
\begin{equation}\label{eqn:result9}
\left\{
\begin{aligned}
H_{n}^{2}=&\;H_{n-1}^{2}\left[g_{n}\left(Z_{R_{n}}+1\right)^{2n+1}+\ldots\right], \\
H_{n-1}^{2}=&\;H_{n-2}^{2}\left[g_{n-1}\left(Z_{R_{n-1}}+1\right)^{2n-1}+\ldots\right], \\
\vdots
\end{aligned}
\right.
\end{equation}
By setting $n = 1$, \eqref{eqn:result9} becomes the standard form for Hubble parameter
\begin{equation}
H_{1}^{2}=H_{0}^{2}\left[g_{1}(Z_{R}+1)^{3}+\ldots\right].
\end{equation}

This model demonstrates that the dissolution of $D_1$ branes and the compactification of dimensions lead to significant changes in the Hubble parameter. In this model, $D_n$ branes are formed from joining $D_1$ branes. During the formation of $D_n$ branes from $D_1$ branes, some amount of energy becomes free. Then, some dimensions are compacted, and some extra energies are produced. These energies cause the expansion of branes and the universe and change the cosmic parameters like the Hubble parameter.

\section{Conclusion}\label{conclusion}
Previously, it has been shown that our universe could live on one of the branes of a BIon and interact with other universes on other branes through a wormhole. This interaction could cause a difference between the number of degrees of freedom on the surface of the universe brane and the number of degrees of freedom in a Bionic bulk. The wormhole in a Bion could be a $D_1$ brane or a $D_2$ brane. By dissolving this $D_1$ brane into the universe of $D_3$ branes of a Bion, they expand. On the other hand, each Dn brane of a BIon can form by joining $n$ D1 branes. By joining $D_1$ branes to each other and forming $D_3$ branes, some amount of energy is produced. Additionally, it is possible that some $D_1$ branes merge with $D_3$ branes and then become compact. During compaction, some energy is released. These energies cause the expansion of the universe and the evolution of cosmic parameters, such as the Hubble constant. The form of the Hubble parameter that is obtained in this model is very similar to the form of the Hubble parameter in cosmological models.


\backmatter

\section*{Declarations}

\begin{itemize}
\item Funding: N/A
\item Conflict of interest/Competing interests: The authors declare that there are no conflicts of interest
\item Ethics approval and consent to participate: N/A
\item Consent for publication: Full consent of all authors is granted for publication
\item Data availability: N/A 
\item Materials availability: N/A
\item Code availability: N/A
\item Author contribution: Both authors have contributed equally to this research
\end{itemize}

\input sn-article.bbl

\end{document}